\documentclass[runningheads]{llncs}
\usepackage{microtype}
\usepackage{multirow}

\usepackage{booktabs} 
\usepackage{caption} 
\usepackage{subcaption} 
\usepackage[all]{nowidow}
\usepackage[utf8]{inputenc}
\usepackage{multicol}
\usepackage{hyperref}

\hypersetup{
hidelinks,
pdftitle={AI-enhanced Auto-correction of Programming Exercises: How Effective is GPT-3.5?},
pdfauthor={Imen Azaiz and Oliver Deckarm and Sven Strickroth},
pdfkeywords={E-Assessement; Personalized Feedback; GPT-3.5; Large Language Model; Programming Education; Formative Assessment}
}

\begin{document}
\title{AI-enhanced Auto-correction of Programming Exercises: How Effective is GPT-3.5?}

\titlerunning{AI-enhanced Auto-correction of Programming Exercises with GPT-3.5}

\author{Imen Azaiz\orcidID{0009-0005-6458-4169} \and
Oliver Deckarm \and
Sven Strickroth\orcidID{0000-0002-9647-300X}}
%
%
\institute{LMU Munich, Germany
\email{\{imen.azaiz,sven.strickroth\}@ifi.lmu.de}}

\setcounter{page}{0}
\thispagestyle{empty}
\noindent {\Large Accepted to the \href{https://online-journals.org/index.php/i-jep}{International Journal of Engineering Pedagogy} (iJEP; eISSN: 2192-4880).\\This is the author's version.}
\newpage

\maketitle

\begin{abstract}
Timely formative feedback is considered as one of the most important drivers for effective learning. Delivering timely and individualized feedback is particularly challenging in large classes in higher education. Recently Large Language Models such as GPT-3 became available to the public that showed promising results on various tasks such as code generation and code explanation. This paper investigates the potential of AI in providing personalized code correction and generating feedback. Based on existing student submissions of two different real-world assignments, the correctness of the AI-aided e-assessment as well as the characteristics such as fault localization, correctness of hints, and code style suggestions of the generated feedback are investigated. The results show that 73\,\% of the submissions were correctly identified as either correct or incorrect. In 59\,\% of these cases, GPT-3.5 also successfully generated effective and high-quality feedback. Additionally, GPT-3.5 exhibited weaknesses in its evaluation, including localization of errors that were not the actual errors, or even hallucinated errors. Implications and potential new usage scenarios are discussed.
\end{abstract}


\section{Introduction}
Learning programming is often experienced as hard by students \cite{Luxton-Reilly2018}.
It involves various aspects like mastering the formal programming language syntax, problem-solving skills, computational thinking, and testing/debugging abilities. Hence, learning programming requires practice, commonly in the form of weekly homework assignments in traditional programming courses. Timely feedback is a key driver of learning \cite{Hattie2007}.
However, introductory courses are frequently characterized by large class sizes with several hundred students, making manual teacher corrections for timely feedback impractical due to resource limitations \cite{SB22}.
Crowdsourcing feedback through peer review has limitations, relying on the motivation and expertise of participating students~\cite{Indriasari2020,Heller2019,Strickroth2023}.
Many e-assessment systems have been developed to (semi-)automatically assess students’ submissions and provide feedback \cite{SS22}. 
However, a common issue is that these systems often focus solely on functional correctness, through employ standard compilers and test cases. Moreover, developing sophisticated tests is very laborious.

Recently, powerful and accessible AI technologies such as GPT-3.5 have emerged, demonstrating their ability to solve programming assignments and pass exams \cite{geng2023chatgpt,tian2023chatgpt,Savelka_2023,FinnieAnsley2023MyAW}. Hence, the idea was born to investigate whether the Large Language Model (LLM) GPT-3.5 can be used to assess students’ solutions and deliver automated feedback \cite{Hellas_2023}. This study explores the advantages of an AI-assisted approach in introductory programming education, where assignment instructions and student solutions are sent to GPT-3.5 to obtain tailored feedback at scale. The main research question is whether GPT-3.5 can provide ``good'' personalized feedback, including code style suggestions, for programming tasks. The following research questions are investigated: 
\begin{itemize}
\item \textbf{RQ1}: What is the quality of GPT-3.5 in determining the submissions’ correctness?
\item \textbf{RQ2}: How can the GPT-3.5-generated feedback be characterized regarding length, suggestion hints, compliance to assignment instruction, code correction, error localization, and personalization?
\item \textbf{RQ3} What are the strengths and weaknesses of GPT-3.5 in delivering personalized feedback and code corrections?
\end{itemize}
By addressing these questions, we contribute to the understanding of how AI can be effectively utilized to provide personalized and accurate support in beginner programming education. Our findings provide insights into implications and potential limitations of the AI-assisted approach, benefiting educators and developers aiming to improve programming instruction and support.
The paper's structure includes a presentation of related research in programming education and e-assessment, a description of our empirical study's methodology, the presentation of study results in the evaluation section, and concludes with a discussion and conclusions.

\section{Related Research}
LLMs such as GPT and ChatGPT became available recently and have already been investigated in the domain of programming to assist developers, educators, and learners to explore capabilities and limitations: Tian et al. explored the potential of ChatGPT as an assistant bot for programmers and found ChatGPT to be effective in tasks such as code generation, program repair, and code summarization \cite{tian2023chatgpt}. Moreover, its summary explanations of incorrect code offer valuable insights into the original intentions of developers. Surameery \& Shakor~\cite{Nigar2023} show that ChatGPT can provide effective debugging assistance by predicting and explaining programming errors. In the context of automatic program repair, ChatGPT has shown to have a similar performance on a standard benchmark set as existing deep-learning program repair techniques and outperforms standard program repair approaches \cite{Sobania2023}.

In the domain of education, one of the pioneering papers used OpenAI Codex model and showed that it can solve CS1 programming exercises with a quality comparable to students, but sometimes struggles with output formatting \cite{Finnie2022}. In a follow up study, Codex outscored most students for CS2 assignments and exams \cite{FinnieAnsley2023MyAW}. Becker et al. \cite{becker2022programming} discuss the rapid advancement of AI-driven code generation tools and emphasize the need for educators to address their potential impact on teaching. Danny et al. \cite{denny2023computing} explore the impact of LLMs' on computing education, summarizing recent studies on their performance and pedagogical influence, covering responsible tool use, societal implications, and readiness for dynamic environments.

LLMs have already been successfully used to generate programming assignments \cite{Sarsa_2022} and to explain syntax errors and compiler error messages in an actionable way that can be used as feedback for students \cite{phung2023generating,leinonen2022using}. Furthermore, LLMs have been used to generate good code explanations \cite{Sarsa_2022} and a study by Macneil et al. \cite{macneil2022experiences} showed that automatic generated code explanations by LLMs in an e-book were appreciated by students . In a study by Leinonen et al. \cite{leinonen2023comparing}, students rated generated code explanations as better than those of their fellow peers. This research underlines that LLMs can be used to support students in learning programming and students do not reject their usage. Common is that all authors argue that instructors need to inspect the generated results.

The capabilities of LLMs for providing feedback was already investigated (e.\,g., \cite{kiesler2023exploring,balse2023investigating}). Kiesler et al. \cite{kiesler2023exploring} investigated ChatGPT’s responses to students seeking assistance with introductory programming tasks. They also assessed ChatGPT's performance in providing formative feedback to students on programming tasks, demonstrating its effectiveness in areas like compilation error detection, correction, and textual explanations. However, they only provided the code written by the student to ChatGPT and not the assignment specification. While doing this research, Balse et al. \cite{balse2023investigating} published a paper investigating the feasibility of GPT-3 to check, critique and suggest changes to code written by students in an online programming exam of an undergraduate Python programming course. They found a high variability in the accuracy (between 57 and 79\,\%) and received incorrect and inconsistent feedback. This research focuses on Java programming assignments, analyzes the performance of GPT-3.5 based on the correctness of the submission, and systematically analyses qualitative aspects of the feedback content such as (correct) fault localization, coding style suggestions, and type of the feedback (contains some code, no code, only code).

In summary, prior research demonstrates the effectiveness of LLMs in assisting developers, learners, and educators in programming (education). However, there is a research gap in assessing GPT-3.5's ability to correct student code submissions and providing feedback while maintaining assignment requirements and characteristics and quality of the generated feedback. LLMs offer an advantage by eliminating the need for test case development compared to traditional e-assessment systems. Our research aims to assess GPT-3.5's ability to provide personalized, assignment-compliant feedback for enhanced auto-correction, ensuring effective and valuable corrections that match the submitted code and adhere to the assignment's specifications.

\section{Methodology and Approach}

Our primary goal is to assess the quality of GPT-3.5-generated feedback for student programming assignment submissions. We aim to ascertain whether GPT-3.5 can efficiently provide automated formative feedback for programming assignments. An empirical study involving two programming assignments selected from an introductory course at Ludwig Maximilian University Munich, Germany was conducted. Approximately 900 students, majoring or minoring in computer science, attended this course, which included ten weekly homework assignments. Students uploaded their solutions to the e-assessment system GATE \cite{SOP11} to receive instant feedback. Additionally, a weekly peer review was conducted for one of the assignments (cf. \cite{Strickroth2023}). Participation in both, the assignments and peer review, was voluntary. Teaching assistants held weekly exercise sessions to discuss homework assignment solutions. We selected a less complex assignment from the third assignment sheet and a more challenging one from the seventh for this study. We randomly sampled approx. 10\,\% of student submissions for these two assignments for an in-depth analysis. For privacy and ethical compliance, we exclusively analyzed submissions from students who explicitly consented to their use for research purposes (695 out of 900 students). There are 33 sampled student submissions for the first selected assignment. The instruction is (translated to English): ``Write a Java application named \textit{SimpleWhileLoop} that prints all odd numbers from 1 to 10 using a \textit{while} loop, and prints `Boom!' (without quotation marks) afterwards.'' The instruction of the second selected assignment is as follows (translated to English, 16 sampled submissions): ``This task refers to the abstract data types \textit{singly linked list} and \textit{queue} introduced in the lecture. Implement the (given) Queue interface based on the specifications (in the interface) for a queue using the \textit{QueueImpl} class, using a singly linked list.'' Additionally, the interface for ``Queue'' was provided as a Java file which specified the methods to implement as well as a description of their semantics as JavaDoc. The \textit{Queue} interface defines several methods, including \textit{void append()}, which adds a value to the end of the queue; \textit{boolean isEmpty()}, which checks if the queue is empty; \textit{void remove()}, which removes the first element from the queue; \textit{int peek()}, which returns the value of the first element in the queue (or the given constant \textit{EMPTY\_VALUE} if the queue is empty); and \textit{int[] toArray()}, which returns an array containing the values of the queue. The students were expected to use a static inner class for the node class.

Submissions were processed by the GPT-3.5 API (davinci-003 model) on June 12, 2023, three times in randomized order, resulting in 99 feedbacks for the first assignment and 48 for the second. We experimented with various prompts to suit our research question and settled on the following prompt structure:
\begin{quote}
ASSIGNMENT INSTRUCTION

Find all kinds of errors, including logical ones, and provide hints for their correction or improvement, including suggestions for code style.

STUDENT SUBMISSION
\end{quote}

To ensure a thorough evaluation, two researchers with extensive expertise in correction and providing feedback for programming assignments meticulously reviewed all submissions and the generated feedback. They assessed the submissions’ correctness (both syntactic and functional) and analyzed 147 generated feedback texts for the \textit{SimpleWhileLoop} and Queue tasks. The feedback analysis covered an analysis of the classification quality (\textbf{RQ1}) and various characteristics, including length, content (code-only, text with code, style hints), hint accuracy, references to the specific submissions, and error detection comprehensiveness (\textbf{RQ2}). The applied methods are described in the respective sections before the results are presented.

\section{Results}
In this section, we present the results of the manual evaluation of submissions and the analysis of GPT-3.5's generated feedback. We first assess the feedback length. Next, we report on the correctness and accuracy of GPT-3.5's assessment of the submissions. Finally, the evaluation of the characteristics of the generated feedback is presented.

We examined 49 submissions in total for our investigation across the two assignments. For the \textit{SimpleWhileLoop} assignment, there are 33 submissions, of which 19 are ``correct'' and 14 are ``incorrect''. For the Queue assignment, there are 16 submissions of which 4 are ``correct'' and 12 are ``incorrect''. ``Correct'' submissions met the assignment requirements accurately, while “incorrect” submissions had syntactic or functional errors, did not fulfill the requirements, or were incomplete (e.\,g., missing files or empty classes with no solution attempt).

\subsection{Length of the Generated Feedback}

The feedback length was determined by tokenizing the feedback string using white space (\verb|"\s+"|) and counting the resulting tokens. The overall median length is 82 words ($\bar{x}=85$). For the \textit{SimpleWhileLoop}, the median word count is 65 words ($\bar{x}=72$, min=21, max=156). For the \textit{Queue} assignment, the median word count is 112 words ($\bar{x}=111$, min=19, max=178). 
Tab.~\ref{tab:wordcountall} shows the number of words for both assignments and for all three runs. According to the Mann-Whitney U-Test the difference between the two assignments is statistically significant ($U=3783$, $p=.000$, two-sided).

\begin{table}[h]
\centering
\centering
\begin{tabular}{c|cccr|cccc|}
\cline{2-9}
 & \multicolumn{4}{c|}{\textbf{SimpleWhileLoop}} & \multicolumn{4}{c|}{\textbf{Queue}} \\ \hline
\multicolumn{1}{|c|}{\textbf{\# words in feedback}} & \multicolumn{1}{c|}{\textbf{1st}} & \multicolumn{1}{c|}{\textbf{2nd}} & \multicolumn{1}{c|}{\textbf{3rd}} & \textbf{OA} & \multicolumn{1}{c|}{\textbf{1st}} & \multicolumn{1}{c|}{\textbf{2nd}} & \multicolumn{1}{c|}{\textbf{3rd}} & \textbf{OA} \\ \hline
\multicolumn{1}{|l|}{\textbf{Mean word count}} & \multicolumn{1}{r|}{72} & \multicolumn{1}{r|}{68} & \multicolumn{1}{r|}{75} & 72 & \multicolumn{1}{r|}{121} & \multicolumn{1}{r|}{103} & \multicolumn{1}{r|}{109} & 111 \\ \hline
\multicolumn{1}{|l|}{\textbf{Median word count}} & \multicolumn{1}{r|}{65} & \multicolumn{1}{r|}{67} & \multicolumn{1}{r|}{65} & 65 & \multicolumn{1}{r|}{120} & \multicolumn{1}{r|}{106} & \multicolumn{1}{r|}{108} & 112 \\ \hline
\multicolumn{1}{|l|}{\textbf{Min word count}} & \multicolumn{1}{r|}{28} & \multicolumn{1}{r|}{21} & \multicolumn{1}{r|}{28} & \multicolumn{1}{r|}{21} & \multicolumn{1}{r|}{44} & \multicolumn{1}{r|}{19} & \multicolumn{1}{r|}{28} & \multicolumn{1}{r|}{19} \\ \hline
\multicolumn{1}{|l|}{\textbf{Max word count}} & \multicolumn{1}{r|}{156} & \multicolumn{1}{r|}{116} & \multicolumn{1}{r|}{156} & \multicolumn{1}{r|}{156} & \multicolumn{1}{r|}{174} & \multicolumn{1}{r|}{178} & \multicolumn{1}{r|}{165} & \multicolumn{1}{r|}{178} \\ \hline
\end{tabular}
\caption{Overview of Word Counts in GPT-3.5's Generated Feedback}
\label{tab:wordcountall}
\end{table}

\subsection{Correctness of the GPT-3.5 Evaluation}
\label{sec:quali}

We assessed the correctness of the feedback generated for the 49 submissions over 3 iterations, for a total of 147 feedback instances. Since GPT-3.5 does not always explicitly label submissions as ``correct'' or ``incorrect'', the researchers made this determination: If the feedback solely contains suggestions without identified errors or functional changes, we classified it as being labeled ``correct'' by GPT-3.5. Conversely, if GPT-3.5 identified errors, proposed modifications that address critical issues, or resembled a sample solution without errors or specific hints, we interpreted it as labeled ``incorrect''. Out of the 147 submissions, 69 incorrect submissions were correctly identified as ``incorrect'' by the GPT-3.5 (\textit{TN}, 47\,\%) and 39 correct submissions were identified as ``correct'' (\textit{TP}, 27\,\%). Additionally, 30 correct submissions were identified as ``incorrect'' by the LLM (\textit{FN}, 20\,\%), and 9 incorrect submissions were identified as ``correct'' (\textit{FP}, 6\,\%). Overall, a total of 108 submissions out of 147 were correctly identified (accuracy: 73\,\%). Tab. \ref{tab:correctness} presents an in-depth evaluation of GPT-3.5’s correctness based on different evaluation metrics for the three runs (1st, 2nd, 3rd) as well as the overall average (OA) for each task and both tasks combined. The metrics include accuracy ((\textit{TP}+\textit{TN})/(\textit{TP}+\textit{TN}+\textit{FP}+\textit{FN})) that presents the proportion of correct classifications made by GPT-3.5, precision (\textit{TP}/(\textit{TP}+\textit{FP})) which represents GPT-3.5’s ability to correctly identify correct submissions and minimize false positives, recall (\textit{TP}/(\textit{TP}+\textit{FN})) which indicates GPT-3.5’s ability to correctly identify correct submissions from the total number of actual correct submissions, specificity (\textit{TN}/(\textit{TN}+\textit{FP})) which reflects GPT-3.5’s ability to correctly identify incorrect submissions and minimize false negatives, and finally the F1 score (2$*$\textit{TP}/(2*\textit{TP}+\textit{FP}+\textit{FN})) \cite{larner2021}. The F1 score (Dice Coefficient) is the harmonic mean of precision and recall, providing a balanced measure of GPT-3.5’s performance. Unlike accuracy alone, the F1 score considers both the true positive rate (recall) and the positive predictive value (precision). It assesses the ability to correctly identify positive instances and avoid false positives, offering a more comprehensive evaluation of GPT-3.5’s effectiveness and robustness.

\begin{table}[htb]
\centering
\begin{tabular}{c|cccc||cccc||l|}
\cline{2-10}
 & \multicolumn{4}{c||}{\textbf{SimpleWhileLoop}} & \multicolumn{4}{c||}{\textbf{Queue}} & \multicolumn{1}{c|}{\multirow{2}{*}{\textbf{OA}}} \\ \cline{1-9}
\multicolumn{1}{|c|}{\textbf{Metric}} & \multicolumn{1}{c|}{\textbf{1st}} & \multicolumn{1}{c|}{\textbf{2nd}} & \multicolumn{1}{c|}{\textbf{3rd}} & \textbf{OA} & \multicolumn{1}{c|}{\textbf{1st}} & \multicolumn{1}{c|}{\textbf{2nd}} & \multicolumn{1}{c|}{\textbf{3rd}} & \textbf{OA} & \multicolumn{1}{c|}{} \\ \hline
\multicolumn{1}{|c|}{Accuracy} & \multicolumn{1}{c|}{75\,\%} & \multicolumn{1}{r|}{72\,\%} & \multicolumn{1}{c|}{57\,\%} & 70\,\% & \multicolumn{1}{c|}{81\,\%} & \multicolumn{1}{r|}{75\,\%} & \multicolumn{1}{c|}{81\,\%} & 81\,\% & 73\,\% \\ \hline
\multicolumn{1}{|c|}{Precision} & \multicolumn{1}{c|}{92\,\%} & \multicolumn{1}{c|}{91\,\%} & \multicolumn{1}{c|}{71\,\%} & 85\,\% & \multicolumn{1}{c|}{67\,\%} & \multicolumn{1}{c|}{50\,\%} & \multicolumn{1}{c|}{67\,\%} & 60\,\% & 80\,\% \\ \hline
\multicolumn{1}{|c|}{Recall} & \multicolumn{1}{c|}{63\,\%} & \multicolumn{1}{c|}{61\,\%} & \multicolumn{1}{c|}{52\,\%} & 58\,\% & \multicolumn{1}{c|}{50\,\%} & \multicolumn{1}{c|}{50\,\%} & \multicolumn{1}{c|}{50\,\%} & 50\,\% & 57\,\% \\ \hline
\multicolumn{1}{|c|}{Specificity} & \multicolumn{1}{c|}{92\,\%} & \multicolumn{1}{c|}{92\,\%} & \multicolumn{1}{c|}{71\,\%} & 86\,\% & \multicolumn{1}{c|}{92\,\%} & \multicolumn{1}{c|}{83\,\%} & \multicolumn{1}{c|}{92\,\%} & 88\,\% & 88\,\% \\ \hline
\multicolumn{1}{|c|}{F1 score} & \multicolumn{1}{c|}{75\,\%} & \multicolumn{1}{c|}{70\,\%} & \multicolumn{1}{c|}{60\,\%} & 69\,\% & \multicolumn{1}{c|}{57\,\%} & \multicolumn{1}{c|}{50\,\%} & \multicolumn{1}{c|}{57\,\%} & 57\,\% & 67\,\% \\ \hline
\end{tabular}
\caption{Evaluation Metrics Comparison of GPT-3.5’s Classification Performance Across the Three Runs for the Two Assignments}
\label{tab:correctness}
\end{table}

For the first assignment (\textit{SimpleWhileLoop}), the results show an overall accuracy of 70\,\%, with an overall precision of 85\,\%. However, the recall values hover around 58\,\%.
The F1 score, which reflects the balance between precision and recall shows a similar pattern, with an overall average of 69\,\%. Although, the specificity values show a wide distribution between 71\,\% and 92\,\% across the runs, with an overall of 86\,\%.
GPT-3.5 shows a slightly better performance overall for the second assignment (\textit{Queue}): The accuracy ranges from 75\,\% to 81\,\% across the runs, with an overall accuracy of 81\,\%. The precision shows some variation but still maintains an overall value of 60\,\%. The recall remains consistent at 50\,\% for all 3 runs, indicating that GPT-3.5 consistently identifies only half of the correct submissions correctly. The specificity ranges from 83\,\% to 92\,\%.

The overall accuracy for both tasks (all submissions, and all six runs) is 73\,\%.  A precision of 80\,\%  and specificity of 88\,\% was achieved by GPT-3.5. However, GPT-3.5 achieved a recall of only 57\,\%. The overall F1 score for GPT-3.5's performance in correcting the two programming assignments is 67\,\%.

\subsection{Evaluation of the Generated Feedback Text}
Only looking at the submission classification quality regarding correctness by GPT-3.5 does not capture the feedback quality. Hence, we qualitatively assessed the feedback content, considering external traits like appearance and superficial content, as well as internal traits, such as content quality (cf. Table \ref{tab:charateristics} and \ref{tab:characteristics2}, \textbf{RQ2}). Internal traits categorize the feedback into three groups: feedback without code (lacking Java programming language keywords or variable/method names), feedback with code, and feedback containing only code. The feedback is identified as personalized if it references parts or (variable) names from the student's submission. If the feedback contains suggestions/corrections and nothing wrong was stated, which means that fixing all errors mentioned and implementing the suggestions results in working code or at least does not break the code further, it is categorized as \textit{only correct corrections/suggestions}. \textit{Partially correct} is assigned when only some feedback parts are correct, while others introduce new issues. Finally, \textit{false correction} indicates that the feedback only contains non-existent errors or suggestions resulting in broken code. These explanations show that correctness does not necessarily imply compliance with the assignment. Feedback was deemed \textit{compliant with the assignment} if the suggestions align with the given instructions. This means that both correct and incorrect corrections can be compliant, and for the feedback to be considered compliant, the code does not have to meet all specifications after implementing the suggestions, which may happen if GPT-3.5 misses some errors. \textit{Completely correct correction} signifies that applying the feedback results in a fully correct submission. Finally, we also tracked whether feedback identified and localized mentioned bugs and if these bugs were present in those locations.

The analysis of erroneous feedback revealed patterns, which were then used to group similar feedback. We categorized submissions into three cases: syntactically and functionally correct (SCFC), syntactically correct but not functionally correct (SCFI, meaning they do not meet task requirements), and syntactically and functionally incorrect (SIFI). We separately analyzed the generated feedback for these three cases to gain a deeper understanding of GPT-3.5's behavior and the feedback effectiveness.

\subsubsection{Analyzing the SimpleWhileLoop Assignment} 

\begin{table}[htb]
\centering
\begin{tabular}{l|ccr|ccr|ccr||cl}
\cline{2-12}
\multicolumn{1}{c|}{\textbf{SimpleWhileLoop}} & \multicolumn{3}{r|}{\textbf{SCFC, n=19}} & \multicolumn{3}{r|}{\textbf{SCFI, n=11}} & \multicolumn{3}{c||}{\textbf{SIFI, n=3}} &\multicolumn{2}{r|}{\textbf{All, n=99}} \\ \hline
\multicolumn{1}{|c|}{\textbf{Characteristics of feedback}} & \multicolumn{1}{r|}{\textbf{1st}} & \multicolumn{1}{r|}{\textbf{2nd}} & \textbf{3rd} & \multicolumn{1}{r|}{\textbf{1st}} & \multicolumn{1}{r|}{\textbf{2nd}} & \textbf{3rd} & \multicolumn{1}{r|}{\textbf{1st}} & \multicolumn{1}{r|}{\textbf{2nd}} & \textbf{3rd}  & \multicolumn{1}{r|}{\textbf{Sum}} & \multicolumn{1}{r|}{\textbf{\,\%}} \\ \hline
\multicolumn{1}{|l|}{Feedback without code} & \multicolumn{1}{r|}{1} & \multicolumn{1}{r|}{1} & 2 & \multicolumn{1}{r|}{1} & \multicolumn{1}{r|}{1} & 2 & \multicolumn{1}{r|}{0} & \multicolumn{1}{r|}{1} & 0 & \multicolumn{1}{r|}{9} & \multicolumn{1}{r|}{9}\\ \hline
\multicolumn{1}{|l|}{Feedback text with code} & \multicolumn{1}{r|}{17} & \multicolumn{1}{r|}{17} & 16 & \multicolumn{1}{r|}{10} & \multicolumn{1}{r|}{9} & 8 & \multicolumn{1}{r|}{3} & \multicolumn{1}{r|}{2} & 2 & \multicolumn{1}{r|}{84} & \multicolumn{1}{r|}{85}\\ \hline
\multicolumn{1}{|l|}{Feedback containing just code} & \multicolumn{1}{r|}{1} & \multicolumn{1}{r|}{1} & 1 & \multicolumn{1}{r|}{0} & \multicolumn{1}{r|}{1} & 1 & \multicolumn{1}{r|}{0} & \multicolumn{1}{r|}{0} & 1 & \multicolumn{1}{r|}{6} & \multicolumn{1}{r|}{6}\\ \hline
\multicolumn{1}{|l|}{Personalized feedback} & \multicolumn{1}{r|}{16} & \multicolumn{1}{r|}{18} & 18 & \multicolumn{1}{r|}{11} & \multicolumn{1}{r|}{11} & 10 & \multicolumn{1}{r|}{3} & \multicolumn{1}{r|}{3} & 3 & \multicolumn{1}{r|}{88} & \multicolumn{1}{r|}{89}\\ \hline
\multicolumn{1}{|l|}{Code style suggestion} & \multicolumn{1}{r|}{16} & \multicolumn{1}{r|}{11} & 11 & \multicolumn{1}{r|}{9} & \multicolumn{1}{r|}{7} & 6 & \multicolumn{1}{r|}{0} & \multicolumn{1}{r|}{0} & 0 & \multicolumn{1}{r|}{60} & \multicolumn{1}{r|}{61}\\ \hline
\multicolumn{1}{|l|}{Only correct correction/suggestion} & \multicolumn{1}{r|}{10} & \multicolumn{1}{r|}{10} & 9 & \multicolumn{1}{r|}{6} & \multicolumn{1}{r|}{9} & 6 & \multicolumn{1}{r|}{2} & \multicolumn{1}{r|}{2} & 3 & \multicolumn{1}{r|}{57} & \multicolumn{1}{r|}{58}\\ \hline
\multicolumn{1}{|l|}{Partially correct correction/suggestion} & \multicolumn{1}{r|}{4} & \multicolumn{1}{r|}{1} & 2 & \multicolumn{1}{r|}{3} & \multicolumn{1}{r|}{1} & 2 & \multicolumn{1}{r|}{0} & \multicolumn{1}{r|}{0} & 0 & \multicolumn{1}{r|}{13} & \multicolumn{1}{r|}{13}\\ \hline
\multicolumn{1}{|l|}{Only false correction/suggestion} & \multicolumn{1}{r|}{5} & \multicolumn{1}{r|}{7} & 8 & \multicolumn{1}{r|}{2} & \multicolumn{1}{r|}{1} & 3 & \multicolumn{1}{r|}{1} & \multicolumn{1}{r|}{1} & 0 & \multicolumn{1}{r|}{28} & \multicolumn{1}{r|}{28}\\ \hline
\multicolumn{1}{|l|}{Compliance with assignment} & \multicolumn{1}{r|}{15} & \multicolumn{1}{r|}{12} & 13 & \multicolumn{1}{r|}{8} & \multicolumn{1}{r|}{10} & 6 & \multicolumn{1}{r|}{1} & \multicolumn{1}{r|}{2} & 3 & \multicolumn{1}{r|}{70} & \multicolumn{1}{r|}{71}\\ \hline
\multicolumn{1}{|l|}{Completely correct correction} & \multicolumn{1}{r|}{4} & \multicolumn{1}{r|}{5} & 9 & \multicolumn{1}{r|}{4} & \multicolumn{1}{r|}{5} & 4 & \multicolumn{1}{r|}{2} & \multicolumn{1}{r|}{2} & 2 & \multicolumn{1}{r|}{37} & \multicolumn{1}{r|}{37}\\ \hline
\multicolumn{1}{|l|}{(Fault) localization} & \multicolumn{1}{r|}{11} & \multicolumn{1}{r|}{5} & 6 & \multicolumn{1}{r|}{8} & \multicolumn{1}{r|}{7} & 7 & \multicolumn{1}{r|}{3} & \multicolumn{1}{r|}{3} & 2 & \multicolumn{1}{r|}{43} & \multicolumn{1}{r|}{43}\\ \hline
\multicolumn{1}{|l|}{(Fault) localization correct} & \multicolumn{1}{r|}{9} & \multicolumn{1}{r|}{5} & 6 & \multicolumn{1}{r|}{1} & \multicolumn{1}{r|}{3} & 4 & \multicolumn{1}{r|}{2} & \multicolumn{1}{r|}{2} & 0 & \multicolumn{1}{r|}{32} & \multicolumn{1}{r|}{32}\\ \hline
\end{tabular}
\caption{Characteristics of the Generated Feedback for the SimpleWhileLoop Assignment (SCFC: syntactically and functionally correct; SCFI: syntactically correct and functionally incorrect; SIFI: syntactically and functionally incorrect)}
\label{tab:charateristics}
\end{table}

Tab.~\ref{tab:charateristics} presents the results of the qualitative evaluation of the feedback generated by GPT-3.5. It includes a total of 99 feedback instances for 33 submissions across 3 runs for the \textit{SimpleWhileLoop} assignment. 
For this assignment, 89\,\% of the generated feedback was personalized, providing specific guidance based on the individual submissions. The remaining feedback consisted of general suggestions or simple statements such as ``no errors were found''. Among all the feedback, 61\,\% included code style suggestions, excluding those related to the SIFI case. Furthermore, 37\,\% of the feedback provided completely correct corrections, accurately identifying and addressing all the issues in the submissions. Fault localization was provided for 43\,\% of all the feedback; however, only 32\,\% of them were correct. We clustered the faulty feedback into patterns, such as, \textit{finding false errors}, \textit{presenting the submitted code as the solution}, \textit{incorrect fault localization}, \textit{ignoring instructions}, and \textit{ignoring subtle requirements}

Starting with the case of SCFC, which refers to correct submissions, GPT-3.5 consistently provided 4 completely correct feedback instances over three runs. The localization for suggestions for improvement was also accurate. Most of these suggestions were related to code style, such as using more descriptive variable names and providing comments with examples.
Interesting was the feedback, where GPT-3.5 said that there are no logical errors in the code and provided a suggestion to enhance the code readability by commenting where nearly every line of the submission was commented (such as \verb|i++; // Increment the loop variable|).
Other suggestions included to use more white space, to use \verb|i+=2| instead of \verb|i=i+2| (or adding 1 twice). Improvement suggestions were not limited to code style recommendations, but also included functional alternatives as a possible solution approach for the problem, e.\,g. ``use an if statement inside the loop to only print the odd numbers'' followed by the presentation of an alternative solution.
Throughout all 3 runs, there were 7 to 9 false negative feedback. Although the submissions were correct, the provided feedback included false suggestions and corrections, indicating a \textit{finding false errors} pattern. For instance, suggestions like using a for loop instead of a while loop or recommending to ``consider using a for loop and printing the range of odd numbers instead of printing individual odd numbers'' that made no sense. These suggestions were not compliant with the assignment requirements or contained incorrect statements, such as ``i is initialized to 0, which is an even number, so the loop never reaches the print statement''. There is also a main pattern in most of the feedback, in \textit{finding false errors} that does not exist, e.\,g. GPT-3.5 suggested incrementing a variable by two instead of adding 1 twice where the code was actually already doing exactly that.
An example where GPT-3.5 did not correctly identify a semantically correct solution printing the odd numbers from 1 to 10 is \verb|while (i < 6) {System.out.println(2*i - 1); i++;}| where GPT-3.5 suggested to change \verb|while (i < 6)| to \verb|while (i <= 10)|. 
The pattern \textit{presenting the submitted code as the solution} occurred three times.

For the SCFI cases, out of 11 submissions, in the first two runs, 10 were correctly identified as incorrect by GPT-3.5, except for one submission where GPT-3.5 incorrectly explicitly said that it is correct in all three runs. This submission was quite remarkable, because all numbers are printed out and, in the second run, GPT-3.5 recognized that only odd numbers should be printed and suggested to introduce a condition variable ``isOdd'' to improve readability. In the last run, three additional false positive cases occurred, where either the statement ``No errors were found'' was provided or a code correction was proposed. All feedback provided was personalized. The only case when a sample solution was suggested without further hints or explanations was for a submission that contains only a class name and an empty while loop body. The quality of the feedback texts remained consistently low across the runs, particularly in terms of being completely correct, with a maximum of 5 out of 11 submissions. The number of feedback instances that demonstrated compliance also decreased from 10 to 6 instances.

In the SIFI case, there are 3 submissions that are syntactically and, therefore, also functionally incorrect. GPT-3.5 correctly classified all 3 submissions as “incorrect" in all three runs. There were no false positives. Personalized feedback was generated for each of these submissions throughout the three runs. No feedback included code style suggestions. In the first run, the feedback exclusively consisted of hints with code sections. However, in the subsequent two runs, the feedback varied and included hints with and without, or solely corrected code. However, out of these submissions, only 6 feedbacks were found to be compliant with the task requirements. Fault localization was provided in 8 feedbacks, but only 4 had their errors correctly localized, along with proper correction suggestions. Examples of feedback with correct fault localization include the following (quote from GPT-3.5):
``// Error: The code does not count from 1 but from 0. // Hint: The while loop should start with the variable 'i' set to 1 instead of 0.''

An overall recurring pattern was, however, that GPT-3.5 failed to distinguish the difference in capitalization of the word ``Boom'' in the print statement, as explicitly stated in the assignment instructions, in all 99 feedback instances. Submissions with ``BOOM'' or ``boom'' were handled exactly the same way by GPT-3.5 (\textit{ignoring instructions} or \textit{ignoring subtle requirements} patterns), and other correction suggestions used the exact same capitalization.

\subsubsection{Analyzing the Queue Assignment} 
The qualitative evaluation of the \textit{Queue} assignment is based on 16 submissions, each processed 3 times by GPT-3.5, resulting in 48 generated answers providing feedback. As before we examine the three different cases separately.
All received feedback was personalized, as they referenced the related submissions, when pointing out errors and suggestions. 28 of the 48 answers contained suggestions to improve the code style. With 8 out of 12 answers (SCFC), 3 out of 6 (SCFI), and 17 out of 30 (SIFI), the feedback comprising code style suggestions is equally distributed through all three cases. However, these suggestions consisted merely of general statements, such as the need for comments throughout the code to explain the logic and proper indentation for better readability.

\begin{table}[h]
\centering
\begin{tabular}{l|ccr|ccr|ccr||cl|}
\cline{2-12}
\multicolumn{1}{c|}{\textbf{Queue}} & \multicolumn{3}{r|}{\textbf{SCFC, n=4}} & \multicolumn{3}{r|}{\textbf{SCFI, n=2}} & \multicolumn{3}{c||}{\textbf{SIFI, n=10}} & \multicolumn{2}{r|}{\textbf{All, n=48}}  \\ \hline
\multicolumn{1}{|c|}{\textbf{Characteristics of  feedback}} & \multicolumn{1}{r|}{\textbf{1st}} & \multicolumn{1}{r|}{\textbf{2nd}} & \textbf{3rd} & \multicolumn{1}{r|}{\textbf{1st}} & \multicolumn{1}{r|}{\textbf{2nd}} & \textbf{3rd} & \multicolumn{1}{r|}{\textbf{1st}} & \multicolumn{1}{r|}{\textbf{2nd}} & \textbf{3rd} & \multicolumn{1}{r|}{\textbf{Sum}} & \multicolumn{1}{r|}{\,\%} \\ \hline
\multicolumn{1}{|l|}{Feedback without code} & \multicolumn{1}{r|}{0} & \multicolumn{1}{r|}{0} & 0 & \multicolumn{1}{r|}{0} & \multicolumn{1}{r|}{0} & 0 & \multicolumn{1}{r|}{0} & \multicolumn{1}{r|}{0} & 0 & \multicolumn{1}{r|}{0} & \multicolumn{1}{r|}{0} \\ \hline
\multicolumn{1}{|l|}{Feedback text with code} & \multicolumn{1}{r|}{4} & \multicolumn{1}{r|}{4} & 4 & \multicolumn{1}{r|}{2} & \multicolumn{1}{r|}{2} & 2 & \multicolumn{1}{r|}{10} & \multicolumn{1}{r|}{8} & 9 & \multicolumn{1}{r|}{45} & \multicolumn{1}{r|}{94}\\ \hline
\multicolumn{1}{|l|}{Feedback containing just code} & \multicolumn{1}{r|}{0} & \multicolumn{1}{r|}{0} & 0 & \multicolumn{1}{r|}{0} & \multicolumn{1}{r|}{0} & 0 & \multicolumn{1}{r|}{0} & \multicolumn{1}{r|}{2} & 1 & \multicolumn{1}{r|}{3} & \multicolumn{1}{r|}{6}\\ \hline
\multicolumn{1}{|l|}{Personalized feedback} & \multicolumn{1}{r|}{4} & \multicolumn{1}{r|}{4} & 4 & \multicolumn{1}{r|}{2} & \multicolumn{1}{r|}{2} & 2 & \multicolumn{1}{r|}{10} & \multicolumn{1}{r|}{10} & 10 & \multicolumn{1}{r|}{48} & \multicolumn{1}{r|}{100} \\ \hline
\multicolumn{1}{|l|}{Code style suggestion} & \multicolumn{1}{r|}{4} & \multicolumn{1}{r|}{2} & 2 & \multicolumn{1}{r|}{1} & \multicolumn{1}{r|}{1} & 1 & \multicolumn{1}{r|}{6} & \multicolumn{1}{r|}{5} & 6 & \multicolumn{1}{r|}{28} & \multicolumn{1}{r|}{58} \\ \hline
\multicolumn{1}{|l|}{Only correct correction/suggestion} & \multicolumn{1}{r|}{1} & \multicolumn{1}{r|}{0} & 0 & \multicolumn{1}{r|}{0} & \multicolumn{1}{r|}{0} & 0 & \multicolumn{1}{r|}{3} & \multicolumn{1}{r|}{5} & 3 & \multicolumn{1}{r|}{12} & \multicolumn{1}{r|}{25}\\ \hline
\multicolumn{1}{|l|}{Partially correct correction/suggestion} & \multicolumn{1}{r|}{0} & \multicolumn{1}{r|}{2} & 1 & \multicolumn{1}{r|}{2} & \multicolumn{1}{r|}{2} & 2 & \multicolumn{1}{r|}{5} & \multicolumn{1}{r|}{4} & 4 & \multicolumn{1}{r|}{22} & \multicolumn{1}{r|}{46}\\ \hline
\multicolumn{1}{|l|}{Only false correction/suggestion} & \multicolumn{1}{r|}{3} & \multicolumn{1}{r|}{2} & 3 & \multicolumn{1}{r|}{0} & \multicolumn{1}{r|}{0} & 0 & \multicolumn{1}{r|}{2} & \multicolumn{1}{r|}{1} & 3 & \multicolumn{1}{r|}{14} & \multicolumn{1}{r|}{29}\\ \hline
\multicolumn{1}{|l|}{Compliance with assignment} & \multicolumn{1}{r|}{2} & \multicolumn{1}{r|}{2} & 2 & \multicolumn{1}{r|}{0} & \multicolumn{1}{r|}{0} & 0 & \multicolumn{1}{r|}{6} & \multicolumn{1}{r|}{6} & 7 & \multicolumn{1}{r|}{25} & \multicolumn{1}{r|}{52}\\ \hline
\multicolumn{1}{|l|}{Completely correct correction} & \multicolumn{1}{r|}{1} & \multicolumn{1}{r|}{0} & 1 & \multicolumn{1}{r|}{0} & \multicolumn{1}{r|}{0} & 0 & \multicolumn{1}{r|}{3} & \multicolumn{1}{r|}{3} & 1 & \multicolumn{1}{r|}{9} & \multicolumn{1}{r|}{19}\\ \hline
\multicolumn{1}{|l|}{(Fault) localization} & \multicolumn{1}{r|}{3} & \multicolumn{1}{r|}{2} & 3 & \multicolumn{1}{r|}{2} & \multicolumn{1}{r|}{2} & 2 & \multicolumn{1}{r|}{7} & \multicolumn{1}{r|}{8} & 6 & \multicolumn{1}{r|}{35} & \multicolumn{1}{r|}{73}\\ \hline
\multicolumn{1}{|l|}{(Fault) localization correct} & \multicolumn{1}{r|}{0} & \multicolumn{1}{r|}{0} & 0 & \multicolumn{1}{r|}{1} & \multicolumn{1}{r|}{1} & 1 & \multicolumn{1}{r|}{6} & \multicolumn{1}{r|}{7} & 3 & \multicolumn{1}{r|}{19} & \multicolumn{1}{r|}{40}\\ \hline
\end{tabular}
\caption{Characteristics of the generated feedback for the Queue assignment}
\label{tab:characteristics2}
\end{table}

There are 4 submissions and 12 feedback instances belonging to the SCFC case (cf. Tab.~\ref{tab:characteristics2}). All feedback was given in the form of text with code, as lists of errors and suggestions. In 50\,\% of the feedback, the student submission was identified incorrectly as an incorrect submission. Noticeable is, that only one feedback is completely correct, but three partially correct and eight simply false. This is accompanied by an error localization in eight submissions with not even one being correct. Only 52\,\% of the feedback complies with the instructions. Furthermore, we observed three patterns present in the feedback. The first and most prominent pattern is evident in all feedback, wherein GPT-3.5 references the code originally found in the student submission and presents it as a solution to a non-existent error. This is closely associated with GPT-3.5, assuming that the submitted code contains an error in the student's code that requires correction. For instance, one piece of feedback states, ``The `remove' method should check to see if the queue is empty before trying to remove elements, otherwise there would be an error.'' while the respective method already performs this exact check. The second pattern is present in the feedback for 3 submissions. Here, GPT-3.5 is simply ignoring the given instructions. It suggests to change names, like renaming \verb|isEmpty()| to \verb|isListEmpty()|, or the return types of methods, in contradiction to the given interface twice. The third pattern is seen in feedback for two submissions where GPT-3.5 mistakenly identifies non-existent code errors. Unlike the first pattern, there's no submitted involved. Instead, GPT-3.5 seems to struggle in understanding the code and attempts to fix what it perceives as issues. For instance, it suggests adding an unnecessary ``else'' statement in the ``peek()'' method, stating: ``Error: \verb|peek()| does not include an else case to return \verb|EMPTY_VALUE|. Hint: In the \verb|peek()| method, add an else statement to return \verb|EMPTY_VALUE| when the queue is empty.'' with GPT-3.5 apparently not recognizing, that an else case is not needed for this method to work:
\begin{verbatim}
    public int peek() {
        if(!this.isEmpty()) {
            return head.getValue();
        }
        return EMPTY_VALUE;
    }
\end{verbatim}

For the SCFI case, involving two submissions and six GPT-3.5 feedback instances, one incorrectly classified a submission as correct (cf. Tab. \ref{tab:characteristics2}), while another provided a mix of text and code. While none of the corrections was entirely correct, none was entirely wrong; they were all partially correct. Each feedback included error localization, which was accurate for half of the submissions (3). However, none of the feedback fully complied with the provided instructions, exhibiting two recurring patterns. The pattern of \textit{finding false errors} (only false correction/suggestion) occurred in two feedbacks, and the incorrect error localization observed in the 2nd and 3rd run of the SCFI case can be linked to this pattern. The pattern \textit{ignoring the instructions} was evident in all six corrections (cf. Tab. \ref{tab:characteristics2}). The first of these two patterns can be traced back to the usage of the ternary condition operator inside a return statement, the second pattern shows up, because all submissions do not use the required singly linked list, but a doubly linked list or the LinkedList from the Java collection framework.

For the third case (SIFI), we examined the 30 feedback instances from GPT-3.5 based on ten student submissions, from which two classified the associated submission incorrectly as correct. Three responses consisted of both text and code, and three further responses consisted of code only. The correction of 11 feedback instances was completely correct, 13 corrections were partially correct and six corrections false. A total of 21 corrections has fault localization included, and 16 of these were correct. Moreover, 19 responses of GPT-3.5 conform to the given instructions. All three patterns are also present in the feedback for this case. Interestingly the \textit{presenting the submitted code as the solution} pattern is present in 9 corrections, but 8 of these corrections have no semantic error but only a syntactical error based on a missing semicolon. The \textit{finding false errors} pattern can be found five times. These always correspond to some code where a connection to some other part of code has to be made such as GPT-3.5 mentioning missing null checks but actually this check is done through a method call. Similar false errors are found for \textit{if-else} conditions or \textit{for} loops in a slightly more complicated fashion. Also, code using the ternary condition operator gets tagged again as incorrect. The feedback for 16 submissions contains the \textit{ignoring instructions} pattern. In 18 submissions an implementation of the singly linked list itself was missing (the node class), with GPT-3.5 not correcting these in 15 cases. Furthermore, one correction ignored the given interface and suggested renaming a method. A new observed pattern unique to the third case is that if a submission had significant errors, including syntax and logic errors leading to exceptions, GPT-3.5 provided code solutions.

\section{Discussion}

\subsection{Quality of GPT-3.5 in Determining the Submissions’ Correctness}
For \textbf{RQ1}, we investigated the quality of GPT-3.5 in determining the correctness. On one hand, the overall accuracy shows that 73\,\% of all classifications made by GPT-3.5 are correct. Also, when GPT-3.5 identifies a submission as correct, it is indeed correct in 80\,\% of the cases (precision). However, the recall value is significantly lower, indicating that GPT-3.5 could only identify 58\,\% of all correct submissions in the data set. This implies that a portion of the correct submissions remain unnoticed, which could be discouraging for students whose submissions are correct but still get evaluated as incorrect. The specificity of 88\,\% highlights GPT-3.5’s ability to correctly identify and classify incorrect submissions. On the other hand, a comparison between the two assignments shows significant differences for the precision (\textit{SimpleWhileLoop}: 8\,\% vs. \textit{Queue}: 60\,\%). A reason for this might also lie in the different frequency of correct submissions in the dataset. Furthermore, there are runs in which the accuracy is as low as 57\,\%, the precision as low as 50\,\%, and the specificity as low as 71\,\% (cf. Tab.~\ref{tab:correctness}). Hence, the quality of GPT-3.5’s feedback seems to be dependent on the specific assignment and also on hidden factors related to the GPT-3.5 model. This point of uncertainty has been highlighted in related studies as well (e.\,g., \cite{Hellas_2023}). The overall accuracy of GPT-3.5 (73\,\%) seems to be in line with comparable research on Python tasks (57 to 79\,\%) \cite{balse2023investigating} and to be comparable with student peer reviews where partial incorrect feedback in approx. 14\,\% of the cases \cite{Heller2019} or an accuracy of 64\,\% \cite{Strickroth2023} have been reported.

\subsection{Characterization of the Generated Feedback by GPT-3.5}
An in-depth analysis of the generated feedback (\textbf{RQ2}) reveals for the \textit{SimpleWhileLoop} assignment that only 37\,\% of the feedback were completely correct (cf. Tab.~\ref{tab:charateristics}) and for the \textit{Queue} 25\,\% (cf. Tab.~\ref{tab:characteristics2}). Here, however, needs to be considered whether is pedagogically better to provide a list of all errors at once or only the major ones first (cf. discussion in \cite{Jeuring2022}). Hence, the \textit{completely correct} characterization might be too strict, and the focus should be put on the \textit{only correct correction/suggestion} and \textit{partially correct correction/suggestion} characterizations. In 28\,\% of the cases in which GPT-3.5 provided hints on hallucinated errors, suggested alternative solution paths such as using a for loop instead of a while loop (which is not compliant with the instruction), or incorporating an \textit{if} condition to test for odd numbers instead of using i+2, where the submissions were already correct. Only 58\,\% of the feedback contained only correct correction/suggestion. From these results, it can be concluded that currently GPT-3.5 does not guarantee reliability in providing completely accurate feedback. It may be important to further investigate how students respond to partially correct feedback or feedback that only provides false suggestions for improvement or what other scenarios are beneficial here such as providing the GPT-3.5 feedback to teaching assistants as a draft for semi-automatic feedback provisioning. This also opens up the debate between formative and summative assessment. For summative assessment an accurate diagnosis is fundamental. Formative assessment, however, should help learners learn and reflect on their submission, so it may be more forgiving if occasional errors or incorrect suggestions are included if they foster reflection and exploration. Still, incorrect feedback should be avoided, if possible, to not confuse novices. For peer reviews, often very short feedback is reported (e.\,g., median 13--14 words \cite{Heller2019,Strickroth2023}), here GPT-3.5 provides significantly longer feedback with a median of 82 words. However, students do not get the advantage of seeing other solution strategies and learning to critique code.

One significant advantage of GPT-3.5’s feedback is that it is personalized in all cases for the \textit{Queue} assignment and in 89\,\% of cases for the \textit{SimpleWhileLoop}. Another advantage of GPT-3.5 is error localization ``for free''. For instance, 53\,\% of errors were successfully localized, and among these localized errors, 35\,\% were correctly identified overall for both assignments. These advantages of personalized feedback and error localization sets GPT-3.5 apart from conventional e-assessment systems that require the laborious task of writing functional tests for all anticipated errors to provide fine grained feedback and error localization. With GPT-3.5, the generation of personalized feedback can be automated, providing a scalable and efficient solution for delivering targeted guidance to students. GPT-3.5’s ability to better identify incorrect submissions as incorrect to identify correct submissions as correct may be the prompt used that specifically asked for finding errors. Hence, prompt engineering or using different prompts for different cases might allow to optimize the situations where the use of GPT-3.5 is most appropriate: In the SCFC (syntactically and functionally correct) case, it would be beneficial to let GPT-3.5 suggest code style improvements, as the evaluation showed that GPT-3.5 provided many good coding style suggestions and alternative solutions. This can stimulate creativity and critical thinking among students. In the SCFI (syntactically correct but functionally incorrect) case, the focus would be on querying GPT-3.5 specifically for semantic errors in the submissions and error localization. Finally, in the SIFI (syntactically and functionally incorrect) case, GPT-3.5 showed the best results compared to the other cases. Hence, it is crucial the prompt GPT-3.5 to be more precise about the types of errors, as it has already demonstrated good performance in fault localization. To better detect the different cases, it might be helpful to combine traditional e-assessment systems (with test cases) with GPT-3.5 to enhance the benefits of both systems.

There are several submissions for which GPT-3.5 provided code-only feedback. It seemed as if this only happened when there were too many errors or no real solution attempted by the students. Here it is questionable whether such a model solution without explanation helps struggling students or might encourage gaming the system approaches (cf. \cite{Baker2009}). It would be interesting to compare this with the feedback provided by real teaching assistants or peer reviews.

\subsection{Strengths and Weaknesses of GPT-3.5 in Delivering Feedback}
To answer \textbf{RQ3}, GPT-3.5’s strengths lie in its ability to provide personalized feedback and suitable code corrections that may enhance the learning experience and helping address coding weaknesses effectively at scale quickly. Particularly interesting is that for the \textit{SimpleWhileLoop} GPT-3.5 did not honor the casing of the string to be printed out but focused on the overall correctness of the algorithm (the more difficult part of the assignment; this was also partly the case for the \textit{Queue} assignment; or submissions where the algorithm was basically correct but contained a syntax error). This aligns with findings of a related study that the LLM Codex struggles with output formatting on code generation \cite{Finnie2022}. Although weaknesses include occasional inaccuracies in error identification (particularly for subtle requirements such as the casing of strings), provisioning of model solutions without explanations, suggesting changes violating the assignment instructions, too generic feedback, and limitations with complex tasks.

GPT-3.5 is just one example of an LLM that was chosen here due to its availability and public API. In the future, there might be different LLMs or even more specialized LLMs for this specific task that show a better performance. Further research is needed.

\subsection{Limitations}
It needs to be noted that GPT is under active development. Hence, future versions might yield different results. Additionally, the response highly depends on the questions asked. Hence, using different questions might also have an impact on the results. We experimented with different (slightly) prompts, and all showed comparable results on our tests. As GPT-3.5 is a LLM and is trained to predict subsequent tokens based on preceding ones, its responses can be different even if the very same question is asked. To address this threat, every submission was sent to GPT-3.5 three times and the results are aggregated. Moreover, the results might vary depending on aspects such as the programming language used, and the description of the task. Also, only two assignments were analyzed. The tasks were selected from a real course and, therefore, allow insights into how GPT-3.5 performs on real input. Finally, the manual classifications used here to categorize the responses from GPT-3.5 might be subjective. To mitigate this threat, the classification was conducted by two researchers with an intensive exchange.

\section{Conclusions and Outlook}
In this paper, we explored how effectively the LLM GPT-3.5 can provide personalized feedback, including code correction and style suggestions, for programming tasks. We evaluated GPT-3.5 using student submissions from two real-world assignments in a first-semester course. The results indicate a 73\,\% accuracy rate in correctly identifying submissions as either correct or incorrect. Furthermore, GPT-3.5 provided adequate feedback in 47,\% of cases. Overall, it performed better in identifying incorrect submissions than in recognizing correct ones. An in-depth analysis revealed GPT-3.5's capability to offer personalized feedback by detecting syntax and functional errors in student submissions and providing suggestions for code improvement. However, GPT-3.5 occasionally made errors or suggested changes that did not align with assignment instructions. Due to identified reliability and consistency issues, we advise caution when using LLMs and particularly GPT-3.5 for fully automatic student feedback on programming tasks at this time. Currently, it is not good enough for summative grading. Nonetheless, the current version can be a valuable tool for teaching assistants in pre-assessing large-scale assignments. It can be used to facilitate quick error identification and generating drafts for personalized feedback, reducing the need for writing unit tests. We plan further research to address current weaknesses and determine optimal use cases for LLMs. This includes exploring how to integrate them in e-assessment systems with traditional tests, allowing for different prompts in syntactically correct submissions. An alternative approach might be to use a LLM to repair syntactically incorrect submissions and then use the repaired version for partial scoring of the functional correctness. Finally, various feedback characteristics, as described in Section \ref{sec:quali}, could be automatically measured to determine whether feedback needs regeneration or optimization with a different prompt and learner data. Future versions of GPT and other LLMs will likely see improvements, expanding their potential applications in education as discussed in this paper. Hence, we plan to compare our results with other LLMs.

\section{Acknowledgments}

This research is part of the project AIM@LMU funded by the German Federal Ministry of Education and Research (BMBF) under the grant number 16DHBKI013. The responsibility for the content of this publication lies with the authors.

The authors thank the students of the lecture \textit{Einführung in die Programmierung} of winter semester 2021/22 who allowed us to use their submissions for this research.

\bibliographystyle{IEEEtran}
\bibliography{mainABP23} 

\end{document}